\documentclass[prl,twocolumn,showpacs,a4paper]{revtex4-1}
\usepackage{amssymb}
\usepackage{amsmath}
\usepackage{graphicx}
\usepackage{hyperref}

\newcommand{\ket}[1]{\left|#1\right\rangle}
\newcommand{\abs}[1]{\left|#1\right|}

\begin{document}

\title{Opto-mechanical transducers for long-distance quantum communication}

\author{K. Stannigel$^{1}$}
\author{P. Rabl$^{2,1}$}
\author{A. S. S\o rensen$^{3}$}
\author{P.\ Zoller$^{1}$}
\author{M.~D.\ Lukin$^{2,4}$}

\affiliation{$^1$Institute for Quantum Optics and Quantum Information, 6020
Innsbruck, Austria, and\\
Institute for Theoretical Physics, University of Innsbruck, 6020 Innsbruck,
Austria}
\affiliation{$^2$Institute for Theoretical Atomic, Molecular and Optical
Physics, Cambridge, MA 02138, USA}
%Anders
\affiliation{$^3$QUANTOP, Niels Bohr Institute, University of Copenhagen, DK-2100 Copenhagen \O, Denmark}
\affiliation{$^4$Physics Department, Harvard University, Cambridge, Massachusetts 02138, USA}

\date{\today}

\begin{abstract}
We describe a new scheme to interconvert stationary and photonic qubits which is based on indirect qubit-light interactions mediated by a mechanical resonator. This approach does not rely on the specific optical response of the qubit and thereby enables optical quantum interfaces for a wide range of solid state spin and charge based systems. We discuss the implementation of state transfer protocols between distant nodes of a quantum network and % evaluate the resulting state transfer fidelities under realistic experimental conditions.
show that high transfer fidelities can be achieved under realistic experimental conditions.
%show that for the specific examples of electronic spin qubits and superconducting charge qubits high fidelity quantum communication protocols can be implemented under realistic experimental conditions. 
% For the specific examples of electronic spin qubits and superconducting charge qubits we show that high fidelity quantum communication protocols can be implemented under realistic experimental conditions. 
\end{abstract}

% Cavity resonators fiber optics, 42.81.Qb 
% Cavity resonators integrated optics, 42.82.Et 
% Cavity resonators optical, 42.79.Gn
%
% Micromechanical devices, 07.10.Cm
%
% quantum noise 42.50.Lc
%
% Quantum communication, 03.67.Hk
% 
% Quantum algorithms and protocols quantum information, 03.67.Ac
% 
% Quantum information, 03.67.-a 
% 	entanglement production, 03.67.Bg 
%	optical implementations, 42.50.Ex 
%	quantum algorithms and protocols, 03.67.Ac
%
% Mechanical effects of light, 37.10.Vz, 42.50.Wk
%\pacs{42.50.Lc,          42.50.Wk,           07.10.Cm           }
%42.50.Lc 	Quantum fluctuations, quantum noise, and quantum jumps 
%07.10.Cm 	Micromechanical devices and systems
\maketitle

Many %potential 
quantum information applications rely on efficient ways to distribute quantum states either within  a large computing architecture or over long distances for quantum communication. For this purpose optical `flying' qubits play a unique role and the ability to interconvert  `stationary' qubits and photons is a key element in quantum computing and quantum communication architectures. Light-matter interfaces and state transfer protocols have been proposed and first implemented with atomic systems using cavity QED~\cite{CiracPRL1997,KimbleNature2008%DuanMonroe2010
}. 
In light of the remarkable progress in nano-engineered solid state quantum systems, 
the challenge is now to develop equivalent optical interfaces for a broad range of solid state spin \cite{SpinQubitReview} and charge \cite{ChargeQubitReview, QDCavity} based qubits. 
A promising avenue towards this goal is provided by opto-nanomechanics~\cite{AnetsbergerNatPhys2009,EichenfieldNature2009,OptoReview}, where a nano-scale mechanical oscillator can be coherently coupled to light. As described below, this provides a natural setting for an opto-mechanical  transducer (OMT), where indirect qubit-photon interactions are mediated by vibrations of a macroscopic mechanical device.  
%compatible with integration in a scalable optical quantum network. 

\begin{figure}
\begin{center}
\includegraphics[width=0.45\textwidth]{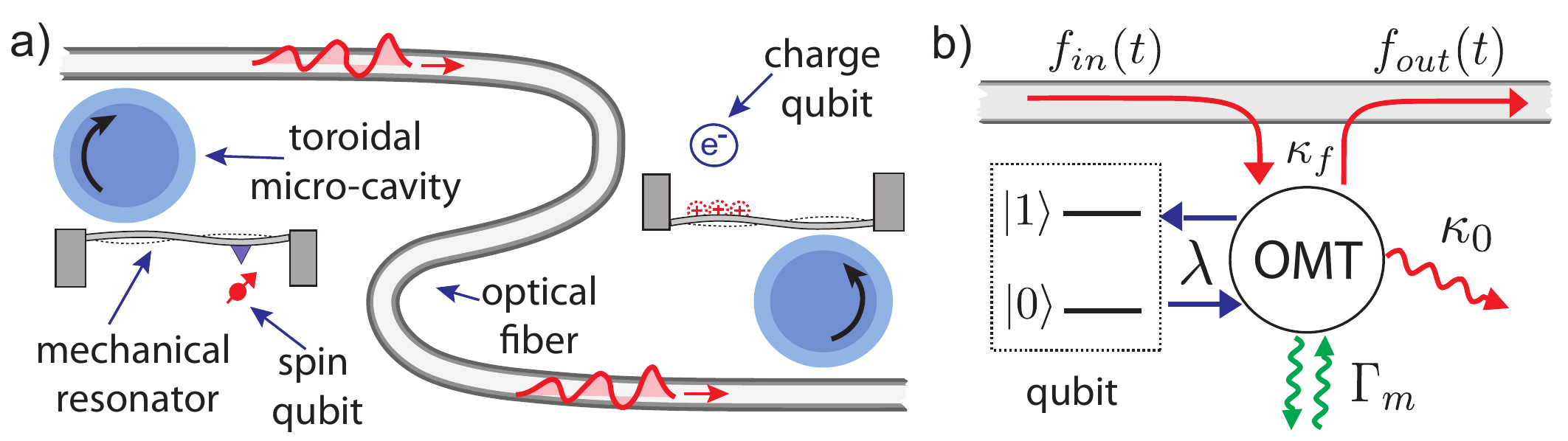}
\caption{a) A quantum network where  spin or charge based qubits and photons are coupled by an opto-mechanical transducer (OMT).
%a) A quantum network where interactions between spin or charge based qubits and photons are mediated by an opto-mechanical transducer (OMT). 
The OMT is realized by a nano-mechanical beam evanescently coupled to a circulating mode of a toroidal microcavity.  b) At each node the OMT mediates coherent coupling between the qubit and photons in the fiber, but also adds noise and loss channels in form of mechanical dissipation $(\Gamma_{th})$ and intrinsic cavity decay $(\kappa_0)$. 
See text for details.
}\label{Figure1}
\end{center} 
\end{figure}

%The basic idea of an OM transducer is shown in Fig.~\ref{Figure1}. 
The setup of Fig.~\ref{Figure1} describes a quantum network where the nodes are represented by solid state qubits and the quantum channel by an optical fiber. %
%
%The basic idea of an OM transducer is shown in Fig.~\ref{Figure1}. 
The qubits are encoded in electronic spin or charge degrees of freedom and coupled to the motion of a mechanical beam via magnetic %field gradients 
\cite{RablPRB2009,MaminNanoLett2009} or electrostatic forces~\cite{CPBResonator,lahaye2009}. At the same time the resonator interacts with the evanescent field of a toroidal micro-cavity as recently demonstrated by Anetsberger  {\it et al.}~\cite{AnetsbergerNatPhys2009}. Excitations from the qubit can be transfered to the mechanical oscillator and then mapped onto a traveling photon in a process which does not rely on optical properties of the qubit and allows the qubit to be spatially separated from the light field. Therefore, this scheme is suited for various solid state spin, charge or superconducting qubits which do not interact coherently with light and provides a basic building block 
for many optical quantum communication applications.

%The setup of Fig.~\ref{Figure1} describes a quantum network where the nodes are represented by solid state qubits and the quantum channel by an optical fiber. 

A fundamental task in optical quantum networks is the implementation of a state transfer protocol $\left(\alpha |0\rangle_i+\beta|1\rangle_i\right)|0\rangle_j \rightarrow |0\rangle_i (\alpha|0\rangle_j+\beta|1\rangle_j)$ between two remote qubits $i$ and $j$. This is achieved by converting the qubit state $|1\rangle_i$ into a photon via the OMT which then propagates along the fiber and is reabsorbed at the second node. 
As first outlined in atomic cavity QED~\cite{CiracPRL1997}, the theory of cascaded quantum systems~\cite{CascadedQS} provides a natural framework to describe these processes and in the case of atomic qubits can be used to identify a set of laser control pulses which achieve a state transfer with unit fidelity. Here we show that OMTs allow us to generalize these ideas for a much broader range of qubits.

%Here we show that OMTs allow us to generalize these ideas and to realize controlled quantum communication operations for a much broader range of qubits.

\emph{Model.} We model the setup shown in Fig.~\ref{Figure1}a) by a Hamiltonian $H= \sum_{i=1}^N H_{\rm node}^i + H_{\rm fib}$, 
where $H^i_{\rm node}$ describes the dynamics of node $i$ and $H_{\rm fib}$ accounts for the coupling between the cavities and the fiber.
Following previous work 
\cite{OptoReview,RablPRB2009,CPBResonator}
we obtain for each node $(\hbar=1)$
\begin{equation}\label{eq:Hnode}
\begin{split}
H_{\rm node}^i= 
&H_q^i + \frac{\lambda}{2} \left(\sigma^i_-b_i^\dag + \sigma^i_+ b_i\right) +  \omega_r b_i^\dag b_i   \\
&+ \Delta_c^i c_i^\dag c_i +(G_i c_i^\dag + G_i^*c_i) (b_i+b_i^\dag)\, ,
\end{split}
\end{equation}
where $\sigma^i_\mu$ are Pauli operators for the qubit $i$, and $b_i$ and $c_i$ the bosonic operators for the resonator and the cavity modes, respectively. In Eq.~\eqref{eq:Hnode}
$H_q^i=\omega^i_q\sigma^i_z/2$ where $\omega^i_q$ is the tunable qubit splitting, $\omega_r$ is the mechanical vibration frequency and $\lambda$ characterizes the strength of the qubit-resonator coupling which
can be of magnetic~\cite{RablPRB2009}  or electrostatic  origin~\cite{CPBResonator}.  The last term in Eq.~\eqref{eq:Hnode} describes the linearized opto-mechanical (OM) interactions for a driven cavity mode~\cite{OptoReview}. Here, $G_i\!=\!\alpha_i g_0$ is the enhanced OM coupling for a mean cavity field amplitude $\alpha_i$ and $g_0\!=\!a_{0}\!\times\! \partial \omega_c/\partial x$ is the shift of the cavity frequency $\omega_c$ associated with the mechanical zero point oscillation $a_0$. 
For each node the coupling $G_i$ and the detuning $\Delta^i_c\!=\!\omega_c\!-\!\omega_L\!-\!2|G_i|^2/\omega_r$ can be controlled by the strength and the frequency $\omega_L$ of local driving fields.
Note that the parallel beam orientation as in Fig. 1 causes negligible scattering between right and left circulating modes~\cite{AnetsbergerNatPhys2009} which allows us to consider 
%in Eq.~\eqref{eq:Hnode}
a single cavity mode only~\cite{Mixing}.

%  where we have already assumed that the cavity is driven by a strong coherent laser field of frequency $\omega_L$ such that the operator $c$ represents fluctuations of the cavity field around the classical mean value $\alpha$. With $g_0=\partial \omega_c/\partial x\times a_0$ denoting the cavity frequency shift per zero point oscillation $a_0$, the enhanced opto-mechanical coupling is $G=\alpha g_0$ and $\Delta_c=\omega_c-\omega_L-2|G|^2/\omega_r$ is the detuning from the bare cavity frequency $\omega_c$.  

We assume that the laser-driven cavity modes couple dominantly to the right propagating field in the fiber,
$f_R(t,z)= \frac{1}{\sqrt{2\pi}}\int_0^\infty   f_\omega  e^{-i \omega (t-z/c)} \, d\omega$, where $[f_\omega,f^\dag_{\omega'}]=\delta(\omega-\omega')$. 
%However, other configurations can be implemented by selectively exciting the appropriate circulating or standing wave modes. 
Then, $
H_{\rm fib}= %\int_0^\infty  d\omega \,  \omega f_\omega^\dag  f_\omega +
%i \sqrt{\frac{\kappa_f}{\pi}}\sum_i  \int_0^\infty  d\omega \left( c_i f^\dag_\omega e^{i(\omega t-z_i/c)} - h.c.\right).
i \sqrt{2\kappa_f}\sum_i( c_i^\dag f_R(t,z_i)\! -\! {\rm H.c.})
$, where $2\kappa_f$  is the decay rate into the fiber and $z_i<z_{i+1}$ are the cavity positions along the fiber. For each node we define in- and out-fields 
$f_{in,i}(t)= f_R(t,z_i+0^-)$ and $f_{out,i}(t)= f_R(t,z_i+0^+)$ and model the resulting dissipative dynamics by quantum Langevin equations 
\begin{equation}\label{eq:QLE}
\dot c_i  = i[ H_{\rm node}^i, c_i] -  \kappa  c_i  - \sqrt{2\kappa_f}  f_{in,i}(t) - \sqrt{2\kappa_0}  f_{0,i}(t), 
\end{equation}
together with the relation $f_{out,i}(t)=f_{in,i}(t)+\sqrt{2\kappa_f}c_i(t)$. For the first cavity, $f_{in,1}(t)$ is a $\delta$-correlated noise operator acting on the vacuum state while the input for the successive cavities is determined by the relation $f_{in,i}(t)=f_{out,i-1}(t-(z_{i}-z_{i-1})/c)$. 
In Eq.\,\eqref{eq:QLE} we have introduced a total decay rate  $\kappa=\kappa_0+\kappa_f$ and the vacuum noise operators $f_{0,i}(t)$ to account for an intrinsic cavity loss rate $\kappa_0$.
We must also include damping of the resonator modes which for a mechanical quality factor $Q_m=\omega_r/\gamma_m$ is described by the Langevin equations
\begin{equation}\label{eq:ThermalNoiseQLE}
\dot b_i=i[H_{\rm node}^i,b_i] - \frac{\gamma_m}{2}b_i - \sqrt{\gamma_m} \xi_i(t) \, .
\end{equation}
Here, $\langle\xi_i^\dagger(t)\xi_j(t')\rangle=N_{th}\delta_{ij}\delta(t-t')$ and for temperatures $T>\hbar \omega_r/k_B$ we identify below $\Gamma_{th}=\gamma_m N_{th}\approx k_BT/\hbar Q_m$ as the relevant mechanical decoherence rate.  

Eqs.~\eqref{eq:Hnode}-\eqref{eq:ThermalNoiseQLE} describe a cascaded quantum network \cite{CascadedQS} where at each node the OM system acts as a \emph{linear} transducer between the fiber in- and out-fields, the qubit state as well as thermal noise  (see Fig. 1b)). In the absence of the qubits mechanical excitations of the OMT are converted into photons in the fiber with a rate $\gamma_{op}\approx{\rm min}\{|G_i|^2\kappa/(\kappa^2+(\Delta_c^i\!-\!\omega_r)^2),\kappa/2\}$. This rate is given by the smallest real part of the eigenvalues of the linear system \eqref{eq:QLE}-\eqref{eq:ThermalNoiseQLE} for $\lambda\rightarrow0$ and is equivalent to the OM cooling rate in the weak and strong coupling regime \cite{ModeSplitting}.
%
%
%For $\lambda\rightarrow0$
%the dynamics of the OMT is characterized by the rate $\gamma_{op}\approx{\rm min}\{|G_i|^2\kappa/(\kappa^2+(\Delta_c^i\!-\!\omega_r)^2),\kappa/2\}$, which is smallest real eigenvalue of the linear system \eqref{eq:QLE}-\eqref{eq:ThermalNoiseQLE}. It can be interpreted as the rate at which mechanical excitations are converted into photons in the fiber and equivalent to the OM cooling rate in the weak and strong coupling regime \cite{ModeSplitting},
% 
%The dissipative dynamics of cavity and resonator at a given node takes place on a scale $\gamma_{op}$, which is the smallest real eigenvalue of the linear system \eqref{eq:QLE}-\eqref{eq:ThermalNoiseQLE} with $\lambda\rightarrow0$. It is given by $\gamma_{op}\approx{\rm min}\{|G_i|^2\kappa/(\kappa^2+(\Delta_c^i\!-\!\omega_r)^2),\kappa/2\}$ and can be identified with the OM cooling rate in the weak and strong coupling regime \cite{ModeSplitting}, giving the rate at which mechanical excitations are converted into photons in the fiber. 
To proceed, we focus on the experimentally relevant regime where $\lambda\ll\gamma_{op}$,
%To proceed, we now 
%focus on the experimentally relevant regime where $\lambda$ is small compared to $\gamma_{op}\approx{\rm min}\{|G_i|^2\kappa/(\kappa^2+(\Delta_c^i\!-\!\omega_r)^2),\kappa/2\}$, which is the slowest dissipative time-scale of the subsystem containing cavities and resonators.
%It is precisely the rate at which mechanical excitations are converted into photons in the fiber and can be identified with the OM cooling rate in the weak and strong coupling regime \cite{ModeSplitting}. 
%focus on the experimentally relevant regime where $\lambda$ is small compared to the OM cooling rate $\gamma_{op}:={\rm min}\{\kappa/2,|G_i|%^2\kappa/(\kappa^2+(\Delta_c^i\!-\!\omega_r)^2)\}$ at which mechanical excitations are converted into photons in the fiber \cite{ModeSplitting}. 
where we can adiabatically eliminate the fast dynamics of the coupled OM degrees of freedom.  As a result we  obtain a master equation for the reduced qubit density operator $\rho$~\cite{LongVersion}, which we display here for the relevant case of two qubits:
%
%%%% OLD
%Eqs.~\eqref{eq:Hnode}-\eqref{eq:ThermalNoiseQLE} describe a cascaded quantum network \cite{CascadedQS} where at each node the OM system acts as a \emph{linear} transducer between the fiber in- and out-fields, the qubit state as well as the thermal noise  (see Fig. 1b)). To proceed, we now 
%focus on the experimentally relevant regime where $\lambda$ is small compared to $\gamma_{op}\approx{\rm min}\{|G_i|^2\kappa/(\kappa^2+(\Delta_c^i\!-\!\omega_r)^2),\kappa/2\}$. This is the rate at which mechanical excitations are converted into photons in the fiber and can be identified with the OM cooling rate in the weak and strong coupling regime \cite{ModeSplitting}. 
%%focus on the experimentally relevant regime where $\lambda$ is small compared to the OM cooling rate $\gamma_{op}:={\rm min}\{\kappa/2,|G_i|%^2\kappa/(\kappa^2+(\Delta_c^i\!-\!\omega_r)^2)\}$ at which mechanical excitations are converted into photons in the fiber \cite{ModeSplitting}. 
%In this limit we can adiabatically eliminate the fast dynamics of the coupled OM degrees of freedom  and obtain a master equation for the reduced qubit density operator $\rho$~\cite{LongVersion}, which we display here for the relevant case of two qubits:
%%%%% END OLD
%
%
\begin{equation}\label{eq:CascadedME}
\begin{split}
\dot \rho\simeq &  -i (H_{\rm eff} \rho -\rho H_{\rm eff}^\dag) + \mathcal{S} \rho \mathcal{S}^\dag 
+\mathcal{L}_{\rm noise}(\rho)\,.
\end{split}
\end{equation}
Here,  %the first two terms represent the dynamics of an ideal cascaded quantum network where 
%\begin{equation}
%$H_{\rm eff}=\sum_i (H_q^i  -\eta\frac{i}{2}\Gamma_i\sigma_+^i\sigma_-^i ) - i \sum_{i<j}J_{ij} \sigma_-^i\sigma_+^j $
$H_{\rm eff}\!=\! \sum_i H^i_q \!-\! \frac{i}{2}J_{12}(\sigma_-^1\sigma_+^2 \!- \!\sigma_+^1\sigma_-^2)\!-\!\frac{i}{2}\mathcal{S}^\dagger \mathcal{S}$
%\end{equation} 
is an effective (non-hermitian) Hamiltonian and the collective jump operator $\mathcal{S}\!=\!\sum_{i}\sqrt{\eta\Gamma_i}\sigma_-^i$ accounts for dissipation due to photons lost through the fiber. Further, $\eta\!=\!\kappa_f/\kappa$ and the decay rates $\Gamma_i\!=\!2 {\rm Re} \{S_{ii}(\omega_q)\}$ as well as the photon mediated qubit-qubit coupling $J_{12}\!=\!\abs{S_{21}(\omega_q)}\!\simeq\! \eta\sqrt{\Gamma_1\Gamma_2}$ are given by the spectrum 
$
S_{ij}(\omega) = \frac{\lambda^2}{4}\int_0^\infty d\tau  \, \langle [b_{i}(\tau), b^\dag_{j}(0)]\rangle_0 \, e^{i\omega \tau}
$, 
and the resonator equilibrium correlation functions  $\langle b_{i}(\tau) b^\dag_{i}(0) \rangle_0$ follow from the linear Langevin equations~\eqref{eq:QLE} and \eqref{eq:ThermalNoiseQLE} in the limit $\lambda\rightarrow 0$.  
%which we use in the following to study the dependence of $\Gamma_i$ and $N_i$ on the underlying OM parameters.
The last term in Eq.~\eqref{eq:CascadedME} summarizes all decoherence processes in the system and can be written as 
$\mathcal{L}_{\rm noise}(\rho)\simeq\frac{1}{2}\sum_i   \Gamma_i N_i  (  [\sigma_-^i,[\rho,\sigma_+^i]] +{\rm H.c.})+\mathcal{L}_{\kappa_0}(\rho)$. 
Here,
$\mathcal{L}_{\kappa_0}(\rho)=(1\!-\!\eta)\sum_i\frac{\Gamma_i}{2}( 2\sigma^i_-\rho\sigma^i_+-\{\sigma^i_+\sigma^i_-,\rho\})$ accounts for photon losses while other noise sources discussed below are described by effective thermal occupation numbers
$
N_i\!=\! (\lambda^2/2\Gamma_i)\! \times\!{\rm Re} \int_0^{\infty}d\tau \langle b^\dag_{i}(\tau) b_{i}(0) \rangle_0 e^{-i\omega_q\tau}
$. Since coherent processes occur on a timescale $\Gamma_i^{-1}$ the  parameters $N_i$ and $(1\!-\!\eta)$ quantify the imperfections of the system. 
Note that in Eq.~\eqref{eq:CascadedME} we have absorbed a small shift of the qubit frequencies into the $\omega_q^i$, and phases $\theta_i$ into the qubit operators, $e^{i\theta_i}\sigma^i_-\!\rightarrow\!\sigma^i_-$, to obtain real $J_{12}$.

%\emph{Discussion.} The first two terms in Eq.~\eqref{eq:CascadedME} represent the dynamics of an ideal cascaded qubit network~\cite{CiracPRL1997,CascadedQS}.  It is convenient to rewrite $H_{\rm eff}$ in a symmetric form $H_{\rm eff}= \sum_i H^i_q - \frac{i}{2}\sum_{i<j} J_{ij}(\sigma_-^i\sigma_+^j - \sigma_+^i\sigma_-^j)-\frac{i}{2}\mathcal{S}^\dagger \mathcal{S}$ such that $J_{ij}$ is a coherent coupling between qubits while the collective jump operator $\mathcal{S}$ accounts for dissipation due to photons lost through the fiber. In both terms the decay rates $\Gamma_i \equiv\Gamma(\lambda,\omega^i_q,G_i,\Delta^i_{c},\kappa)$ are 

\begin{figure}
\begin{center}
\includegraphics[width=0.45\textwidth]{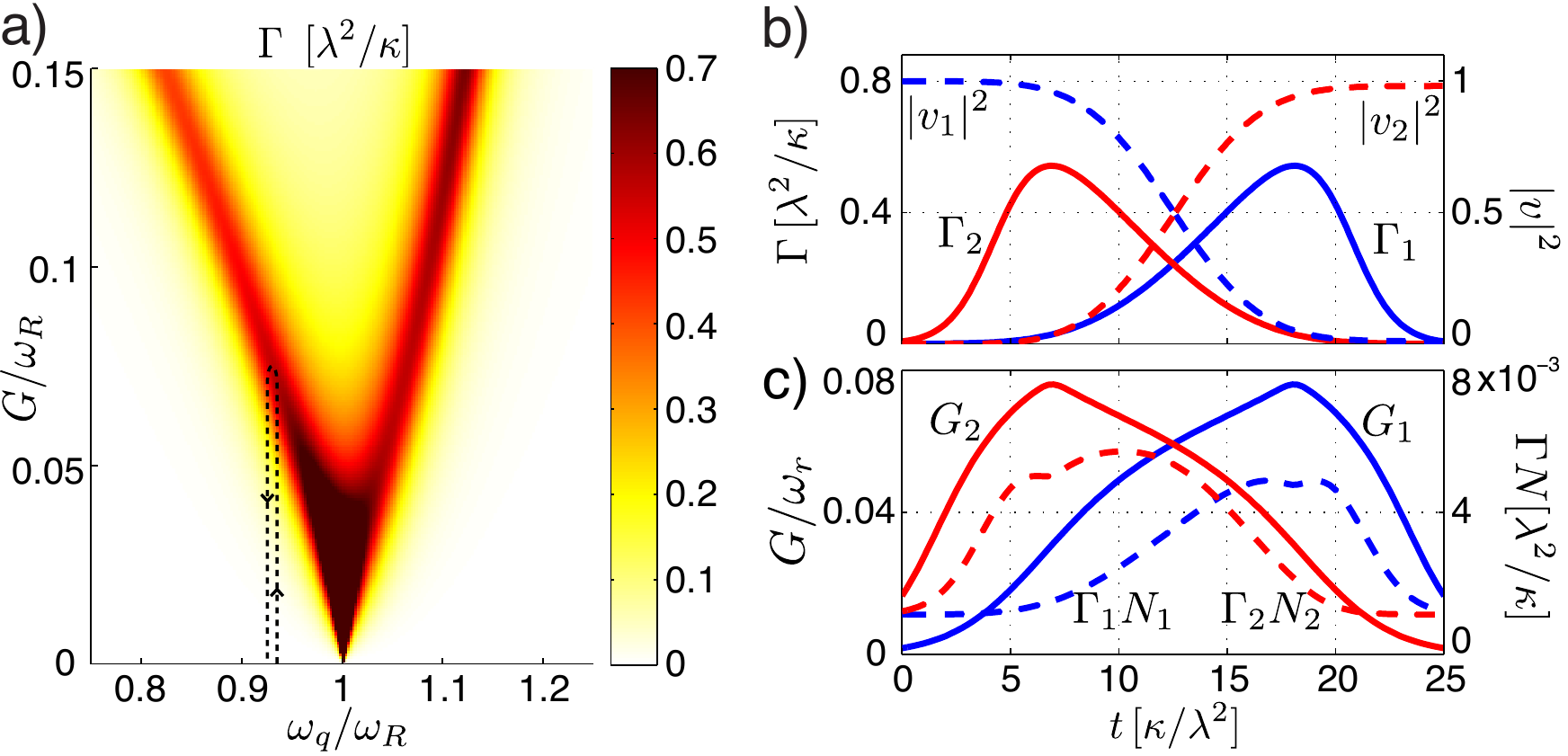}
\caption{a) Effective single-qubit decay $\Gamma$ as a function of $G$ and $\omega_q$ for the parameters  $\kappa_f=0.05 \omega_R$, $\kappa_0=\gamma_{m}=0$, with cavity and resonator being in resonance ($\Delta_c=\omega_R$) at $G=1.5\kappa_f$. The dotted line indicates the control pulse shown in (c).
b) Pulse shapes for $\Gamma_{1,2}(t)$ which implement a perfect state transfer $v_2(t_f)\simeq 1$ as described in the text. c) Control pulses for $G_{1,2}(t)$ which generate the $\Gamma_{1,2}(t)$ shown in (b). The dashed lines indicate the corresponding noise terms which appear in $\mathcal{L}_{\rm noise}$. The parameters used for this plot are $\omega_q^{1,2}=\omega_R-1.5 \kappa_f$, $\Gamma_{th}/\kappa_f=0.01$ and all others as in (a).}
\label{Figure2}
\end{center} 
\end{figure}

\emph{Discussion.} The first two terms in Eq.~\eqref{eq:CascadedME} represent the dynamics of an ideal cascaded qubit network~\cite{CiracPRL1997,CascadedQS}.
%Note that the hermitian and non-hermitian parts of $H_{\rm eff}$ interfere to produce the uni-directional interaction $J\sigma_-^1\sigma_+^2$. 
The coherent and incoherent dynamics of the system is fully determined by the effective decay rates $\Gamma_i$, % \equiv\Gamma(\lambda,\omega^i_q,G_i,\Delta^i_{c},  \kappa)$
 which for $\gamma_m\ll\gamma_{op}$ can be approximated by 
%dropping the terms  $c^\dagger b^\dagger$ and $c b$ in Eq.\,\eqref{eq:Hnode} (rotating-wave approximation) and by assuming $\gamma_m\ll\gamma_{op}$:
%determined by the OM system and they are 
\begin{equation}\label{eq:decayRate}
\Gamma_i\simeq \frac{\lambda^2}{2} \frac{\kappa  |G_i|^2}{(|G_i|^2 + (\Delta_c^i-\omega_q^i)(\omega_q^i-\omega_r))^2+ \kappa^2(\omega_q^i-\omega_r)^2}.
\end{equation}
For $\Delta_c^i\approx\omega_r$ exact values for $\Gamma_i$ are plotted in Fig.~\ref{Figure2}a) as a function of $|G_i|$ and $\omega_q$. 
Its behavior reflects the excitation spectrum of the coupled OM modes at the qubit frequency $\omega_q$.  For  $|G_i|< \kappa/2$ we have a single resonance at $\omega_q=\omega_r$ and a width $\gamma_{op} \simeq |G_i|^2/\kappa$. %This corresponds to the case where the qubit excitation is first transferred 
%to the resonator and successively decays into the cavity. 
For larger $|G_i|$ a mode splitting occurs and two resonances appear at $\omega_\pm\simeq \sqrt{\omega_r^2\pm 2|G_i|\omega_r}$~\cite{ModeSplitting}. This splitting indicates a hybridization of the mechanical and optical mode which then both decay with a rate $\gamma_{op}\simeq \kappa/2$.   

By adiabatically adjusting different OM parameters the qubit decay rate can be tuned within a wide range $\Gamma_{\rm res} \lesssim \Gamma_i(t)\lesssim  \lambda^2/(2\gamma_{op}) < \kappa$, with a small residual decay $\Gamma_{\rm res}\ll\gamma_m$ due to mechanical damping  \cite{LongVersion}.
%
%
%%%% Original
%Therefore, this setting is formally analogous to the cavity QED setup discussed in Ref.~\cite{CiracPRL1997}. Similar arguments can now be applied to identify optimal control pulses for the implementation of state transfer protocols. 
%
%%%% Shortened version
Hence, this setup is analogous to the cavity QED setting of Ref.~\cite{CiracPRL1997} and similar arguments lead to optimal control pulses for state transfer protocols.
%
% 
%First, we can follow the general arguments outlined above to identify ideal pulse shapes for $\Gamma_i(t)$ and second relate those pulses to different experimental control parameters $G_i(t)$, $\Delta_c^i(t)$, $\omega_q^i(t)$ via the explicit expression for the $\Gamma_i$.  
%First, we can follow the general arguments outlined above to identify ideal pulse shapes for $\Gamma_i(t)$ and second use Eq.~\eqref{eq:decayRate} to relate those pulses to different experimental control parameters $G_i(t)$, $\Delta_c^i(t)$, $\omega_q^i(t)$, etc.  
We illustrate this 
for two nodes with $z_1<z_2$ and write the two qubit wavefunction as   
$\ket{\psi(t)}=\alpha \ket{00}+\beta(v_1(t)\ket{10}+v_2(t)\ket{01})$. Initially,  $v_1(0)=1$ and $v_2(0)=0$ 
and our goal is to find pulse shapes for $\Gamma_{1,2}(t)$ which achieve $v_1(t_f)=0$ and $v_2(t_f)=1$ at some final time $t_f$.
A necessary condition for a perfect state transfer is that the system remains in a pure qubit state, $\rho(t)=|\psi(t)\rangle\langle \psi(t)|$. For the ideal case  ($\mathcal{L}_{\rm noise}\equiv 0$, $\eta=1$) this is achieved when the dark state condition $\mathcal{S}(t)|\psi(t)\rangle=(\sqrt{\Gamma_1(t)}\sigma_-^1\!+\!\sqrt{\Gamma_2(t)}\sigma_-^2)|\psi(t)\rangle=0 $ is fulfilled $\forall t$. Together with the Schr\"odinger equation $\partial_t\ket{\psi}=-i H_{\rm eff}(t)|\psi\rangle$ this requirement leads to a set of differential equations which in a first step we can use to derive a set of ideal decay rates $\Gamma_{1,2}(t)$. 
In a second step, Eq.~\eqref{eq:decayRate} allows us to relate those pulses to the actual experimental control parameters $G_i(t)$, $\Delta_c^i(t)$, $\omega_q^i(t)$, etc.

As a specific example we plot in Fig.~\ref{Figure2}b) the time-symmetric pulse $\Gamma_2(t)=\Gamma_1(-t)$,
where $\Gamma_1(\tilde t\!=\!t\!-\!t_f/2)\!=\!\Gamma_0\exp(- c \tilde{t}^{\,2})/(1-\! \Gamma_0 \sqrt{\pi/4c} \,\textrm{Erf}(\sqrt{c}\,\tilde{t}\,))$, 
together with the resulting evolution of  $|v_{1,2}(t)|^2$. Here, $\Gamma_0=\Gamma_1(t=t_f/2)$  and $c>\pi\Gamma^2_0 /4$ are used to adjust the pulse such that $|v_1(t_f)|^2<10^{-2}$.  Fig.~\ref{Figure2}c) shows the corresponding control pulses $G_{1,2}(t)$ which can be used to actually implement the transfer protocol by adjusting the driving strength for each cavity. Alternatively, we can use Eq.~\eqref{eq:decayRate} to identify a similar control pulse for  $\Delta_{c}^{1,2}(t)$ and vary the cavity frequencies $\omega_c^i(t)$~\cite{TunableToroids}. In both cases the mutual dependence of $G_i$ and $\Delta_c^i$ must be taken into account and 
tuning the qubit frequencies ensures that $\delta(t)\equiv\omega_q^2(t)-\omega_q^1(t)+\dot\theta_{1}(t)-\dot\theta_2(t)=0$, $\forall t$.  

\emph{Noise.} 
%Under realistic conditions the OMT adds noise to the system ($\mathcal{L}_{\rm noise}\!\neq\!0$). Since the state-transfer takes place at a rate $\Gamma_i$, the gate error is proportional to $N_i\approx N_{0,i}+N_{{\rm casc},i}$, where
%$N_{{\rm casc},i}$ is defined below and $N_{0,i}$ accounts for noise generated locally by each OM system, 
Under realistic conditions the OMT adds noise to the system which is characterized by $N_i \approx N_{0,i}+N_{{\rm casc},i}$.
Here, $N_{{\rm casc},i}$ is defined below and $N_{0,i}$ accounts for noise which is generated locally by each OM system, 
\begin{equation}\label{eq:N0}
N_{0,i}\approx \frac{\Gamma_{th}}{2\kappa} \frac{\kappa^2+(\Delta_c^i-\omega^i_q)^2}{|G_i|^2}+ \frac{\kappa^2+(\Delta_c^i-\omega^i_q)^2}{4\Delta_c^i\omega^i_q}.
\end{equation}
%The first contribution $\sim \Gamma_{th}$ arises from thermal excitations of the mechanical mode while the second term in Eq.~\eqref{eq:N0} is associated with Stokes scattering events due to energy non-conserving terms as $G_i b_i^\dag c_i^\dag$ in $H^i_{\rm node}$. 
The contribution $\sim \Gamma_{th}$ arises from thermal excitations of the mechanical mode while the second term results from Stokes scattering events due to energy non-conserving terms as $G_i b_i^\dag c_i^\dag$ in $H^i_{\rm node}$. 
On resonance, i.e. $\Delta_c^i=\omega_r$ and $\omega_q\simeq \omega_\pm$, Eq.~\eqref{eq:N0} is similar (but not identical) to the final occupation number in OM cooling experiments~\cite{CoolingTheory,GroundStateCoolingExp}. 
Therefore,  the requirements for ground state cooling, namely $\Gamma_{th}/\gamma_{op} \ll 1$ and sideband resolved conditions $G,\kappa \ll \omega_r$,  are, in addition to $1-\eta\ll1$, also sufficient to realize a low noise OM transducer with $N_0\ll 1$.

Noise photons generated at one node can propagate along the fiber and affect successive nodes, which is described by $N_{{\rm casc},i}$. This contribution  is absent for the first cavity and approximately given by $N_{{\rm casc},i}(\omega_q)\!\approx\! 2\kappa_f\sum_{j<i} \langle c_{j}^\dag(\omega_q)c_{j}(\omega_q)\rangle_0$ otherwise, where $c_i(\omega)$ is the Fourier representation of $c_i(t)$.  For two nodes this  leads to a small asymmetry between $N_1(t)$ and $N_2(t)$ as shown in Fig.~\ref{Figure2}c), but in a larger network the scaling $N_{{\rm casc},i}\!\sim\! (i\!-\!1)\!\times\! N_0$ can limit the number of \emph{active} nodes. This problem can be avoided  by activating individual nodes selectively and one possible scheme to achieve this is outlined below.  Finally, we point out that for $\Gamma_{th}\!\rightarrow\!0$ noise is dominated by Stokes scattering.
%$\sim G_i b^\dag_ic^\dag_i$ in $H_{\rm node}$. 
Since the $G_i$ in different nodes are phase coherent this leads to non-trivial noise correlation effects which have not been included in the approximate form of $\mathcal{L}_{\rm noise}$ given above. 
However, in the sideband resolved regime these effects only lead to corrections of order $G_i^2/\omega_r^2$, $\kappa^2/\omega_r^2$
and they are fully taken into account in the following numerical simulations.

\begin{figure}
\begin{center}
\includegraphics[width=0.45\textwidth]{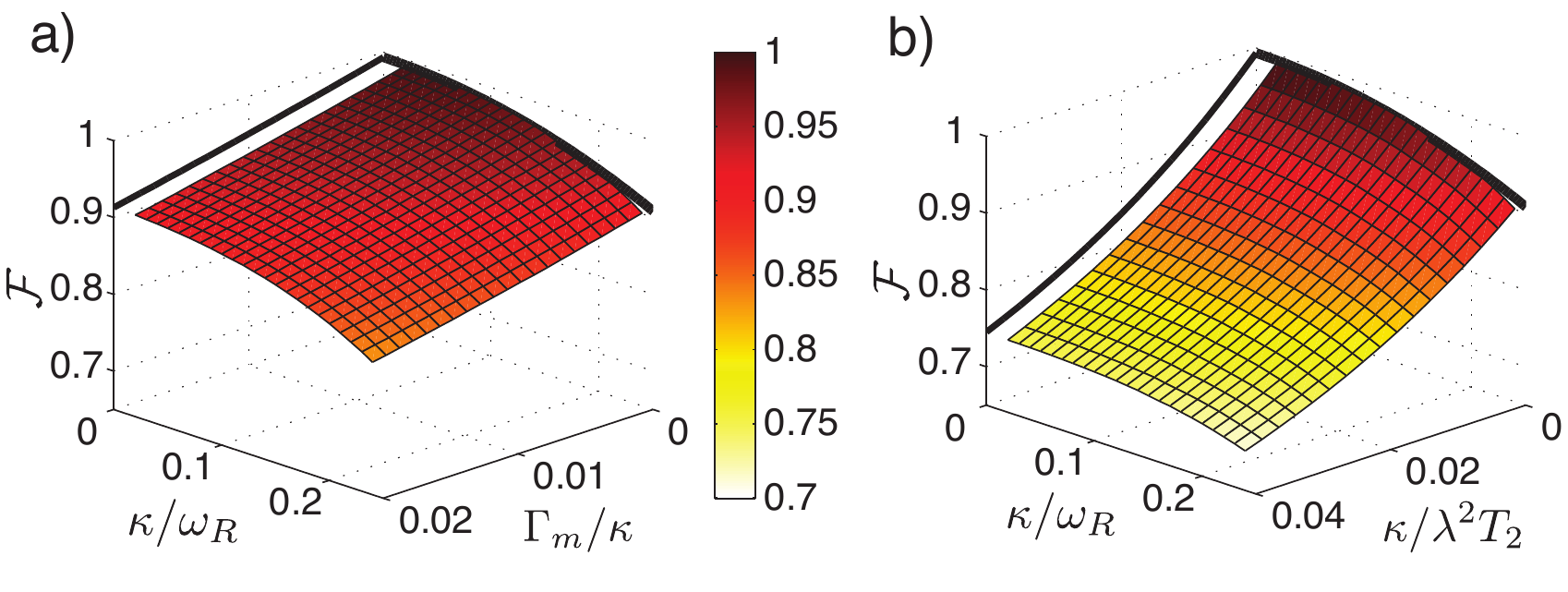}
\caption{a) State transfer fidelity obtained from a numerical simulation of Eq.~\eqref{eq:CascadedME} for the control pulses shown in Fig.~\ref{Figure2} and $\kappa_0=0$. b) The same plot for $\Gamma_{th}=0$ but including an exponential loss of qubit coherence $\sim e^{-t/T_2}$ during the transfer. From these two plots we extract the numerical coefficients $\mathcal{C}_1\approx 4$, $\mathcal{C}_2\approx 1.4$ and $\mathcal{C}_3\approx 7.5$, which appear in the approximate expression of $\mathcal{F}$ given in Eq.~\eqref{eq:Fidelity}. } 
\label{Figure3}
\end{center} 
\end{figure}

To study the quantum state transfer $|\psi_0\rangle_1|0\rangle_2\rightarrow |0\rangle_1|\psi_0\rangle_2$ under realistic conditions we numerically simulate the full master equation~\eqref{eq:CascadedME} for the control pulses described in Fig.~\ref{Figure2}.  The resulting state transfer fidelity $\mathcal{F}=\langle \psi_0|\mathrm{Tr}_1\{\rho(t_f)\}|\psi_0\rangle$ averaged over all input states $|\psi_0\rangle$ is plotted in Fig.~\ref{Figure3}a) for an ideal qubit and in Fig.~\ref{Figure3}b) for qubits with a finite dephasing time $T_2$. 
For small infidelities the results can be summarized as
\begin{equation}\label{eq:Fidelity}
\mathcal{F}\approx 1 - \frac{2}{3}\frac{\kappa_0}{\kappa}-\mathcal{C}_1 \frac{\Gamma_{th}}{\kappa}  - \mathcal{C}_2 \frac{\kappa^2}{\omega_r^2}- \mathcal{C}_3\frac{\kappa }{\lambda^2 T_2} \, ,
\end{equation}
where individual errors arise from intrinsic cavity losses, mechanical noise, Stokes scattering and the qubit dephasing, respectively. The numerical coefficients $\mathcal{C}_i\sim \mathcal{O}(1)$ (see Fig.~\ref{Figure3}) depend on the specific control pulse  and can be optimized for a given set of experimental parameters. 

\emph{Example.} We consider a micro-toroidal cavity with a diameter $d=20\,\mu$m coupled to a doubly clamped SiN beam of dimensions $(l,w,t)\approx  (15,0.05, 0.05)\,\mu$m.
Optical quality factors of $Q_c \ge 2\times 10^9$ \cite{OpticalQ}  correspond to $\kappa_0/2\pi \le 50$ kHz and $\kappa_f/2\pi\simeq1-5$ MHz can be adjusted by the cavity-fiber separation. Depending on the tensile stress the first excited mechanical mode has a frequency  of  $\omega_r/2\pi\approx 5-50$ MHz and a zero point motion $a_0\approx (1.6-0.6)\times 10^{-13}$ m respectively. At $T=100$ mK a mechanical quality factor of $Q_m\sim 2\times 10^5$ corresponds to $\Gamma_{th}/2\pi\sim 10$ kHz and for these parameters the conditions $\Gamma_{th},\kappa_0\ll\kappa\ll \omega_r$ for a high quality OMT are satisfied. 
For electronic spin qubits dephasing times approaching $T_2\approx 10$ ms have been demonstrated~\cite{SpinCoherence}
and following Ref.~\cite{RablPRB2009} we estimate a magnetic coupling strength of $\lambda/2\pi \approx 50$ kHz.
% and for spins dephasing times approaching $T_2\approx 10$ ms have been demonstrated~\cite{SpinCoherence}.
For superconducting charge qubits the electrostatic coupling can be substantially stronger, $\lambda/2\pi\approx 5$ MHz~\cite{lahaye2009}, while in current experiments $T_2=2$ $\mu$s~\cite{ChargeQubitDecoherence}. 
The effective qubit splitting $\omega_q\sim\omega_r$ required for the control pulse described in Fig.~\ref{Figure2} can be engineered using microwave-assisted qubit-resonator coupling schemes~\cite{RablPRB2009, CPBResonator}.
%Note that $\omega_q$ can be an effective splitting of a driven system \cite{RablPRB2009, CPBResonator}, such that we may assume $\omega_q\sim\omega_r$.
By choosing $(\kappa,\omega_r)=2\pi\times(1,5)$ MHz for the spin and $(\kappa,\omega_r)=2\pi\times(5,50)$ MHz for the charge qubit 
we obtain in both cases $\mathcal{F}\approx 0.85$. This shows that state transfer fidelities $\mathcal{F}>2/3$ required for quantum communication \cite{MasserPRL1995} can be achieved with present technology. Near unit fidelities $\mathcal{F}\simeq 0.95-0.99$ can be expected from further optimizations of the system design and control pulses, or from communication protocols which, e.g., correct for photon loss errors~\cite{EnkPRL1997}.

\begin{figure}
\begin{center}
\includegraphics[width=0.45\textwidth]{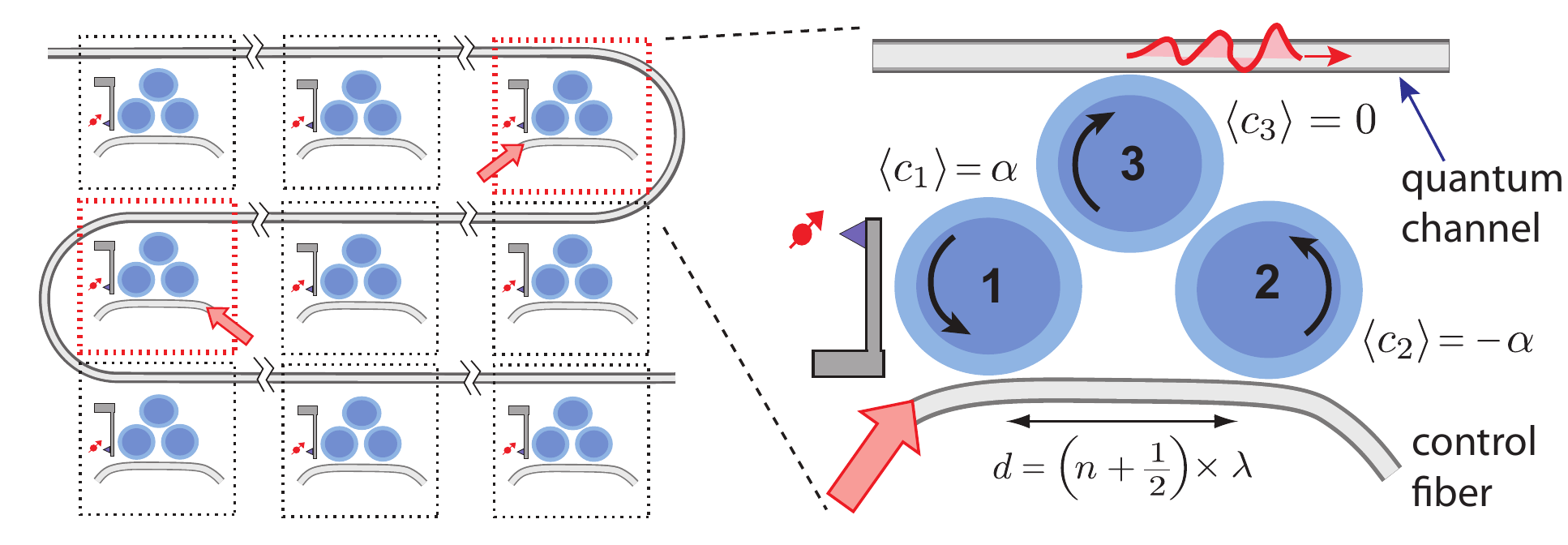}
\caption{Scalable quantum networks based on multi-mode OM transducers. At each node the three cavities are modeled by $H_c=\sum_{n=1}^{3} \Delta_n c_n^\dag c_n +J(c_s^\dag c_3+{\rm H.c.})$ where $J$ is the tunneling coupling of mode $c_3$ with $c_s=(c_1+c_2)/\sqrt{2}$. The driving field applied through the `control fiber' excites the asymmetric mode $c_a=(c_1-c_2)/\sqrt{2}$ such that for $\Delta_1=\Delta_2$ we obtain $\langle c_1\rangle=-\langle c_2\rangle=\alpha$ and $\langle c_3\rangle=0$. The motion of the resonator modulates $\Delta_1$ and induces a linearized OM coupling as given in Eq.~\eqref{eq:Hnode} with the empty mode $c= (c_s+c_3)/\sqrt{2}$. Thereby,  laser noise from the control fields does not affect the quantum channel and nodes can be selectively activated. The direction of the driving field is used to send photons into different directions to connect any two nodes of a large network.
}
\label{Figure4}
\end{center} 
\end{figure}

In conclusion, we have described a universal approach for coherent light-matter interfaces based on OM transducers.
% and discussed their potential applications for optical quantum communication.  
In Fig.~\ref{Figure4} we outline the concept of a multi-mode OM transducer using interference to separate the control fields from the quantum channel. This 
%leads to a suppression of laser noise effects, but also  
enables selective activation of individual nodes to realize large scale solid state or hybrid quantum networks.  
Beyond quantum communication, various other applications for OMTs can be considered such as  new approaches  to engineer single photon non-linearites as well as optical readout and quantum measurement applications for optically non-active quantum systems.

\emph{Acknowledgments}. We gratefully acknowledge discussions with Klemens Hammerer. This work is supported by ITAMP, NSF, CUA, DARPA, the Packard Foundation, and the Danish National Research Foundation. K. S. and P. Z. acknowledge support by the Austrian Science Fund through SFB FOQUS.


\begin{thebibliography}{99}
\bibitem{Mixing} This assumes that $h/|i\Delta_c^i+\kappa|\ll1$ where $h$ is the intrinsic coupling between circulating cavity modes.
\end{thebibliography}
\end{document}